\documentclass[twocolumn,showpacs,preprintnumbers,amsmath,amssymb,floatfix]{revtex4}

\usepackage{graphicx}
\usepackage{dcolumn}
\usepackage{bm}
\usepackage[usenames]{color}
\usepackage{latexsym}

\begin{document}

\preprint{
Growth Law of Bunch Size in Step Bunching Induced by Flow in Solution
}
\title{
Growth Law of Bunch Size in Step Bunching Induced by Flow in Solution
}

\author{Masahide Sato}
\affiliation{
Information Media Center, Kanazawa University,
Kakuma-cho, Kanazawa 920-1192, Japan
}

\date{\today}
\begin{abstract}
{
By carrying out Monte Carlo simulations,
we study step bunching during solution growth.}
For simplicity, we consider a square lattice,
which represents a  diffusion field in a solution,
and  express the diffusion of 
atoms as the hopping of atoms on the  lattice sites.
In our model, we neglect the fluctuation along steps.
An array of steps is expressed as dots on a one-dimensional 
vicinal face.
Step bunching occurs in the case of step-down flow.
{
In previous study (M. Sato: J. Phys. Soc. Jpn. {\bf 79} (2010) 064606),
we studied  step bunching with a slow flow
and showed that 
the width of the fluctuation of step distance increases as $t^\alpha$
with $\alpha =1/3$ in the initial stage.
In this paper, we carry out simulations with a faster flow.
With a faster flow,
the width of the fluctuation of step distance 
first increases as $t^\alpha$ with $\alpha=1/2$,
which is larger than the exponent with a slow flow,
and then an interval during which the width  increases as $t^{1/3}$ appears.
}
\end{abstract}


\maketitle

\section{Introduction}\label{sec:intro}

On a vicinal face, which consists of an equidistant train of straight 
steps, the steps show two types of instability.
One is step bunching, which is the instability of step distance,
and the other is step wandering, which is the instability along steps.
These instabilities are often observed during growth and melting.

During solution growth,
step wandering is caused by the flow in a solution. 
According to a linear stability analysis~\cite{potapenko96jcg},
step wandering is induced by a step-up flow,
in agreement with 
experimental results~\cite{Robey-p00jcg}.
The flow in a solution also causes step bunching.
{
Chernov and coworkers~\cite{Chernov92jcg,Chernov-cm93jcg,Coriell-MCM96jcg} 
assumed that the step density is sufficiently high
that the vicinal face can be treated as a linear sink of atoms,
and they studied the step bunching caused by the flow in a solution.
}
Linear stability analyses and numerical 
studies~\cite{Chernov92jcg,Chernov-cm93jcg,Coriell-MCM96jcg}
have shown that a growing vicinal face is unstable in the case of  a long-wavelength
fluctuation owing to a step-down flow.
The time evolution of an unstable vicinal face was studied by
Bredikhin and Malshakova~\cite{Bredikhin-m07jcg}.
{
According to numerical simulations based on  a nonlinear equation,
a quasi-regular array of high step density,
which is similar to that of bunches observed in 
experiments~\cite{Bredikhin-m07jcg,Bredikhin-GKKM00jcg},
is formed on an unstabilized vicinal face.
}

In previous  studies~\cite{Chernov-cm93jcg,Bredikhin-m07jcg,Chernov92jcg,Coriell-MCM96jcg},
the modulation of step density during step bunching was investigated
but the motion of each step has not been sufficiently studied.
{
Recently, we made a simple discrete model 
and,   by carrying out a Monte Carlo simulation~\cite{Sato-10jpsj}, 
studied the motion of each step during bunching.
In the model, we assumed that the steps are straight
and we expressed the steps as points on a one-dimensional vicinal face.
To represent the diffusion field in a solution,
we introduced a square lattice and treated 
the diffusion in a solution as the hopping of atoms on the lattice.
In the paper~\cite{Sato-10jpsj},
we carried out a Monte Carlo simulation
and showed that the vicinal face is unstable in the case of a long-wavelength fluctuation.
The width of fluctuation of the step distance, $w$, increases 
as $t^\alpha$ with $\alpha=1/3$.
In the simulation, we used low flow rates.
If the flow rate is changed,
the time evolution of the fluctuation of step distance 
and the behavior of steps during bunching may be changed.
}

{
In this paper, we carry out Monte Carlo simulations with a flow in a solution 
whose flow rate is higher than that in previous paper~\cite{Sato-10jpsj}
and study how the time evolution of bunching changes.
}
In \S\ref{sec:mc}, we introduce our simplified model.
In \S\ref{sec:results}, we report the results of Monte Carlo simulations.
The width of the fluctuation of step distance increases with time
in the initial stage and then becomes saturated.
The growth law of the width of the fluctuation and 
the dependence of the saturated width on supersaturation
are studied. 
In \S\ref{sec:summary}, 
we summarize the results and give a brief discussion.

\section{Model}\label{sec:mc}
\begin{figure}[htp]
\centerline{
\includegraphics[width=0.90\linewidth]{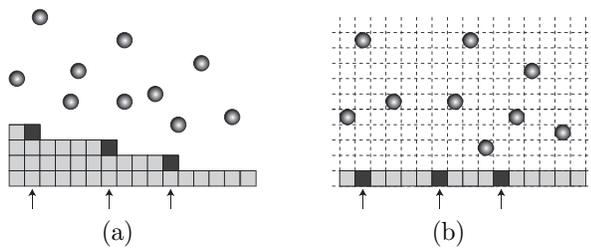}
}
\caption{
Schematic figures of (a) a vicinal face and (b) our model.
Squares and circles represent solid atoms and atoms in solution,
respectively. Dark squares are the solid atoms that form steps.
Arrows show the positions of steps.
}
\label{fig:vicinal face}
\end{figure}

Using a very simplified model (Fig.~\ref{fig:vicinal face})
we carry out Monte Carlo simulations.
In the model, 
a square lattice in a solution is considered.
The diffusion of atoms in a solution is 
expressed as hopping  on the lattice sites. 
The height of steps is neglected,
a vicinal face is expressed as a line,
and steps are treated as points on the line.
The $x$-direction is the step-down direction
and $z=0$ is the altitude of the vicinal face.

In our model, the active atoms are solid atoms forming steps and 
atoms in solution.
In each Monte Carlo trial, we randomly choose one active atom.
When an atom in a solution is chosen, a diffusion trial is carried out.
{
The flow in a solution is expressed as 
the asymmetry of the diffusion probability~\cite{Sato-us98prl,Sato-us00prb}.
When a flow is prarallel to the $x$-direction,
an atom at site $(i,j)$ moves to site $(i, j\pm 1)$ with probability $1/4$ 
and to the site $(i\pm 1, j)$
with probability, $p_\pm= (1 \pm p_\mathrm{flow})/4 $.
The parameter $p_\mathrm{flow}$ represents the effect of the flow
in the solution.
With increasing parameter $p_\mathrm{flow}$, 
the effect of the flow increases.
}
The time is increased after each diffusion trial.
In order that the diffusion coefficient $D$  is equal to $1$,
the increment of the time after each diffusion trial is 
given by $\Delta t= 1/4N_\mathrm{s}$, 
where $N_\mathrm{s}$ is the number of atoms in the solution.

To take account of the  viscosity in the solution,
we assume that $p_\mathrm{flow}$ is proportional to the altitude
from the vicinal face.
When the $z$-coordinate is given by $z=j$, 
$p_\mathrm{flow}$ is expressed  as  
$p_\mathrm{flow}= p_\mathrm{flow}^0 (j-1)$.
Thus, near the vicinal face,
the effect of the flow in the diffusion trial vanishes.
The sign of $p_\mathrm{flow}^0$ shows the direction of the flow.
When $p_\mathrm{flow}^0$ is positive, the flow is in the step-down direction,
and the flow is in the step-up direction for negative $p_\mathrm{flow}$.
If the altitude is sufficiently high that $p_\mathrm{flow}>1$($p_\mathrm{flow}<-1$) , 
we set the hopping probabilities $p_+$ and $p_-$ to be $1(0)$ and $0(1)$,
respectively.
{
In general, we should consider a flow obeys that boundary-layer theory.
However, according to  the Blasius's solution,
the velocity of a smooth flow near a flat surface 
is proportional to the distance from the surface.
Thus, if the effect of the flow far from the surface is neglected,
our model is valid as a simplified model.
}

\begin{figure}[htp]
\centerline{
\includegraphics[width=0.90\linewidth]{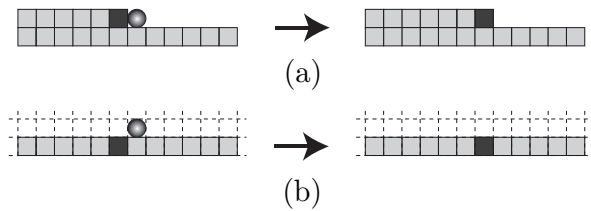}
}
\caption{
Schematic {
 figures} of solidification process in (a) a vicinal face
and (b) our model.
}
\label{fig:solidification}
\end{figure}

\begin{figure}[htp]
\centerline{
\includegraphics[width=0.9\linewidth]{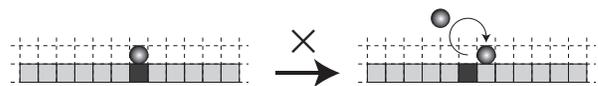}
}
\caption{
Forbidden melting process in our algorithm.
Only one atom can occupy one site. 
}
\label{fig:melting_non}
\end{figure}

The solidification probability $p_+$ and the melting probability $p_-$ 
are given by~\cite{Saito-u94prb}
	\begin{equation}
	p_\pm   = \frac{1}{e^{\mp \phi/k_\mathrm{B}T}+1 },
	\end{equation}
where $\phi$ is the chemical potential gain upon solidification.
A solidification trial is carried out after a diffusion trial.
If the position of the atom is $(i_\mathrm{f},1)$
and that of the step is  $x=i_\mathrm{f}-1$,
the step advances to $x= i_\mathrm{f}$ and the atom in the solution 
vanishes after solidification (Fig.~\ref{fig:solidification}).
\begin{figure}[htp]
\centerline{
\includegraphics[width=0.9\linewidth]{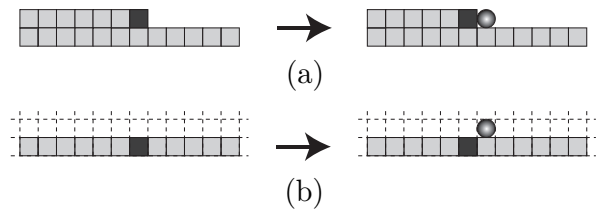}
}
\caption{
Schematic {figures} of melting process in (a) a vicinal face
and (b) our model.
}
\label{fig:melting}
\end{figure}
{
A melting trial is carried out when the chosen active atom is 
an atom forming a step.
If the melting of a step atom occurs at a step edge $x=i_\mathrm{i}$,
}
the step recedes to $x=i_\mathrm{i}-1$ and an atom is expelled to
site $(i_\mathrm{i}, 1)$ (Fig.~\ref{fig:melting}).
In our algorithm, only one atom can occupy a site.
Then, if site $(i_\mathrm{i}, 1)$ is already occupied
by an atom, the melting trial is not carried out (Fig.~\ref{fig:melting_non}).

We assume that the kinetic coefficient at steps is sufficiently large that
the detailed balance stands up at step positions. 
Namely, the concentration of atom is in equilibrium at steps.
The equilibrium concentration in a solution, $c_\mathrm{eq}$,
satisfies
	\begin{equation}
		c_\mathrm{eq} p_\mathrm{s} =(1-c_\mathrm{eq})p_\mathrm{m}.
	\label{eq:detail valance}
	\end{equation}
From eq.~(\ref{eq:detail valance}),
the equilibrium adatom density is given by~\cite{Saito-u94prb}
	\begin{equation}
	 c_\mathrm{eq} = \frac{1}{e^{\phi/k_\mathrm{B}T} +1}.
	\end{equation}
In our model,
when the altitude is larger than the critical value $z_\mathrm{max}$,
the concentration of atoms becomes a constant value  $c_\infty$. 
If $c_\infty$ is higher than $c_\mathrm{eq}^0$,
solidification frequently occurs and the steps advance.
In the opposite case, melting occurs more frequently than solidification 
and the steps recede.

\section{Results of Simulation}\label{sec:results}

We previously studied the motion of steps during step bunching
using the same model~\cite{Sato-10jpsj}.
{
In the previous paper, we showed that a step-up 
flow stabilizes the vicinal face
and that the step bunching occurs with a step-down flow.
Thus, hereafter, we investigate the time evolution of steps
with a step-down flow.}
In the previous study~\cite{Sato-10jpsj}, 
{
we studied the motion of steps with low flow rates.
In an early stage of simulation, a vicinal face seems to be 
unstable for fluctuations with a long wavelength.
The step density gradually modulates
and small bunches are formed. 
When step bunching is caused by the drift of adatoms~\cite{Sato-u99ss},
the formation of bunches changes with the velocity of the drift flow.
A similar change may be generated in the present model.
In the previous study~\cite{Sato-10jpsj},
the flow rate is sufficiently slow that the width of the region 
with a shear flow is $10-100$ lattices.
In the present case, we consider that the shear rate is sufficiently large 
that the width of the region is narrower  than 10.
We carry out simulations with a fast flow
and study how the growth law changes.
}
First, we study the time evolution of the width of the fluctuation 
of step distance, $w$,  
which is defined as 
	\begin{equation}
	w=\sum_{j=1}^{N}
		\sqrt{\frac{1}{n_\mathrm{s}} 
		\sum_{i=1}^{n_s} [l_j(i)-\bar{l}]^2}.
	\end{equation}
$N$ is the number of samples, $n_\mathrm{s}$ is the number of steps, 
$l_j(i)$ is the width of the $i$th terrace in the $j$th sample,
and $\bar{l}$ is the average terrace width.

\begin{figure}[htp]
\centerline{ 
\includegraphics[width=0.5\linewidth]{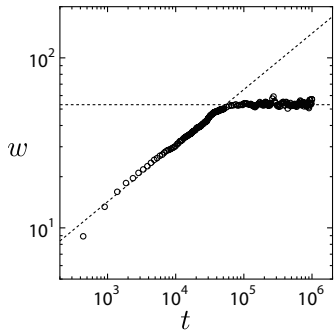}
}
\caption{
Typical time evolution of width of fluctuation of step distance.
The parameters are $\phi/k_\mathrm{B}T=3.0$, 
$p_\mathrm{flow}=0.3$, and $c_\infty=0.2$.}
\label{fig:time evolution of width long}
\end{figure}

Figure~\ref{fig:time evolution of width long} shows 
a typical time evolution of the width of the fluctuation of step distance.
We used the parameters $\phi/k_\mathrm{B}T=3.0$,
$p_\mathrm{flow}=0.3$, and  $c_\infty=0.2$.
The system sizes in the $x$-direction and $z$-direction 
are given by $L_x=512$ and $L_z=128$, respectively. 
Since $n_\mathrm{s}=16$, the average step distance is $\bar{l}=32$.
The data is averaged over $100$ samples.
In the initial stage, $w$ increases with time.
Then, the saturation of $w$ occurs.
Although the time evolution seems to be similar to that with a slow flow,
we study the time evolution in detail.

\begin{figure}[htp]
\centerline{ 
\includegraphics[width=0.45\linewidth]{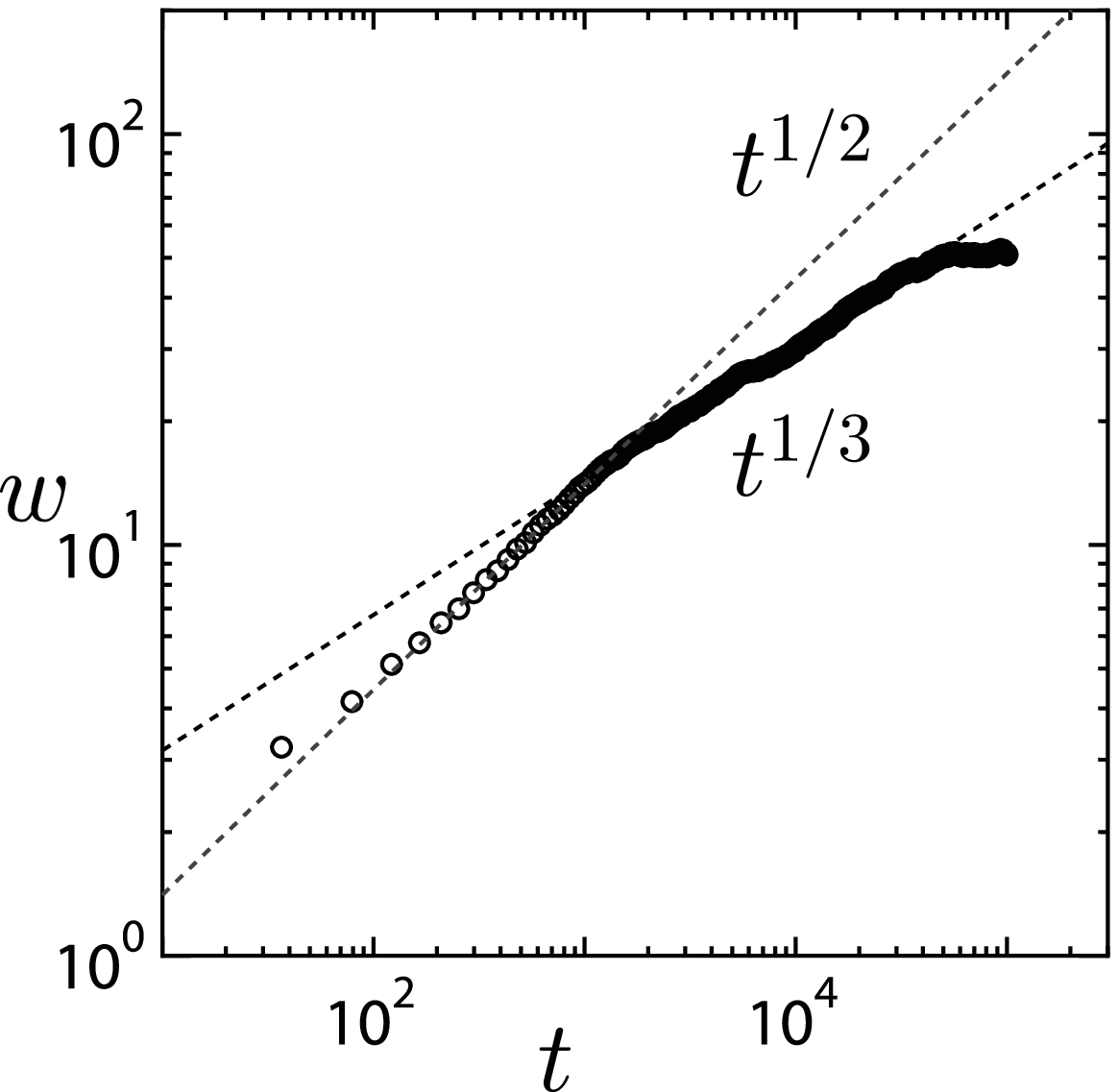}
\includegraphics[width=0.45\linewidth]{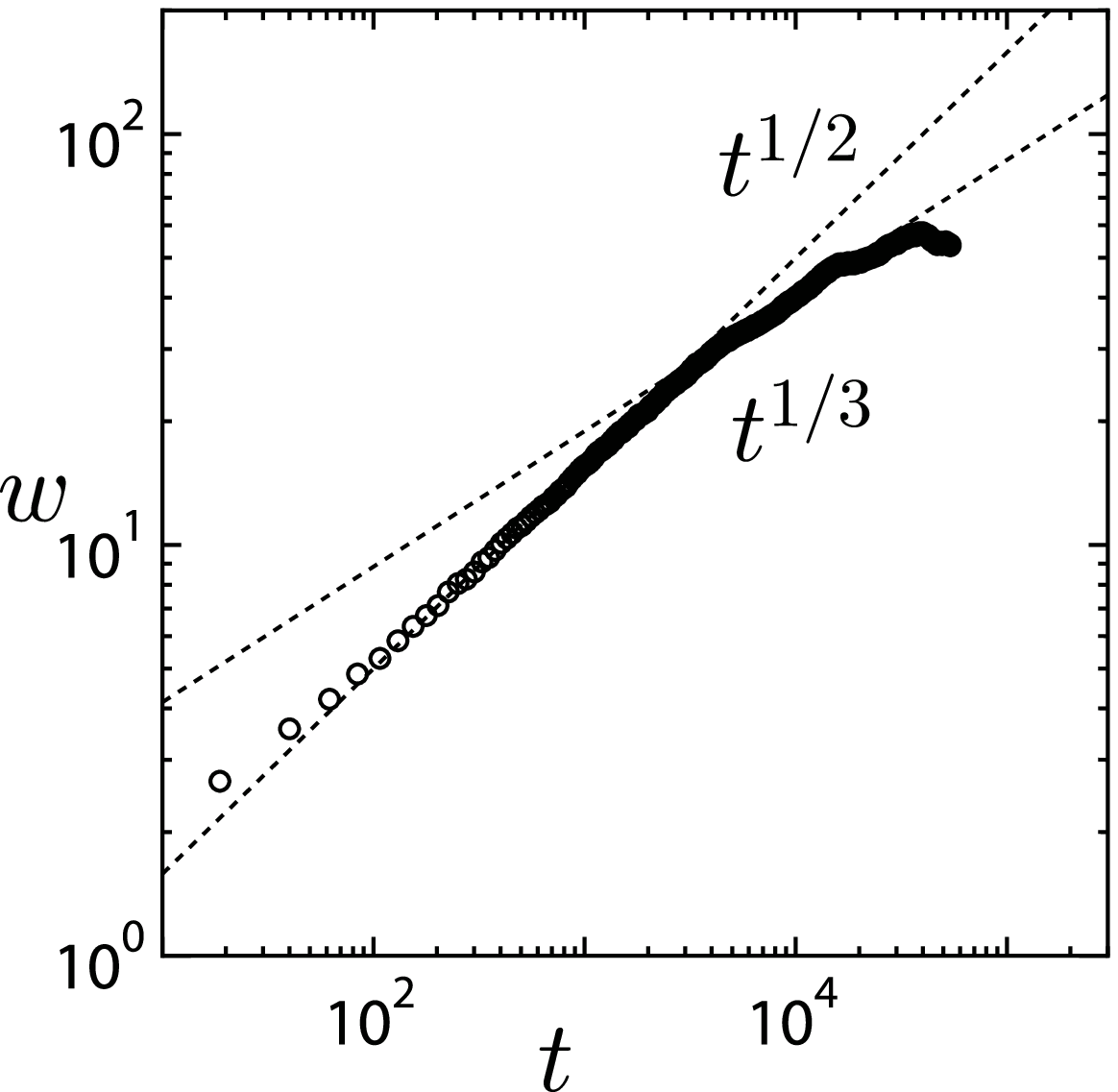}
}
\centerline{\Large (a) \hspace*{3.5cm} (b)}
\caption{
Initial stage of 
time evolution of width of step distance with (a) $c_\infty=0.2$ and 
(b) $c_\infty=0.4$. }
\label{fig:time evolution of width}
\end{figure}

\begin{figure}[htp]
\centerline{ 
\includegraphics[width=0.5\linewidth]{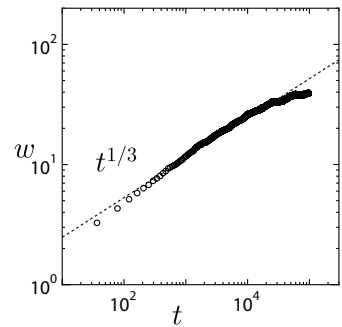}
}
\caption{
Initial stage of 
time evolution of width of step distance with 
$p_\mathrm{flow} = 0.005$.
The other  parameters are the same as those in 
Fig.~\ref{fig:time evolution of width}(a). 
}
\label{fig:time evolution of width long slow}
\end{figure}

Figure~~\ref{fig:time evolution of width} shows 
the initial stage of the time evolution.
The density of atoms far from the vicinal face is given by
$c_\infty=0.2$  in Fig.~\ref{fig:time evolution of width}(a)
and $c_\infty=0.4$ in Fig.~\ref{fig:time evolution of width}(b).
The other parameters are the same as those 
in Fig.~\ref{fig:time evolution of width long}.
As shown in Fig.~\ref{fig:time evolution of width},
the exponent changes with two stages.
In the first  stage, the exponent is given by $\alpha=1/2$,
which is larger than that with a slow flow. 
In the second stage, the growth of $w$ becomes slow and the 
exponent is given by $\alpha=1/3$, which is the same as that
with a slow flow. 
{
In previous study~\cite{Sato-10jpsj},
we studied the time evolution for various slow flows.
In Fig.~\ref{fig:time evolution of width long slow},
we show the time evolution with $p_\mathrm{flow}^0=0.005$
as a sample of the time evolution with a slow flow.
In an early stage ($10^2 \le t \le  10^4$), 
no change in the exponent occurs and $\alpha=1/3$.
}

%
The time at which the exponent $\alpha$ changes from $1/2$
to $1/3$ is about $t \approx 10^3$ in 
Fig.~\ref{fig:time evolution of width}(a)
and $t \approx 3 \times 10^3$   in Fig.~\ref{fig:time evolution of width}(b).
{
The time interval with $\alpha=1/2$ 
in Fig.~\ref{fig:time evolution of width}(a)
is {shorter} than that
in Fig.~\ref{fig:time evolution of width}(b).
This difference is probably caused by the 
increase in the supersaturation. }
However, the time at which the saturation of width occurs
hardly changes irrespective of the supersaturation.
{
Thus, the decrease of time interval with $\alpha=1/3$ 
is probably caused by the decrease in the supersaturation.}
If we use a larger  $c_\infty$,
the time interval in which the exponent is $\alpha =1/3$ may vanish.

\begin{figure}[htp]
\centerline{ 
\includegraphics[width=0.45\linewidth]{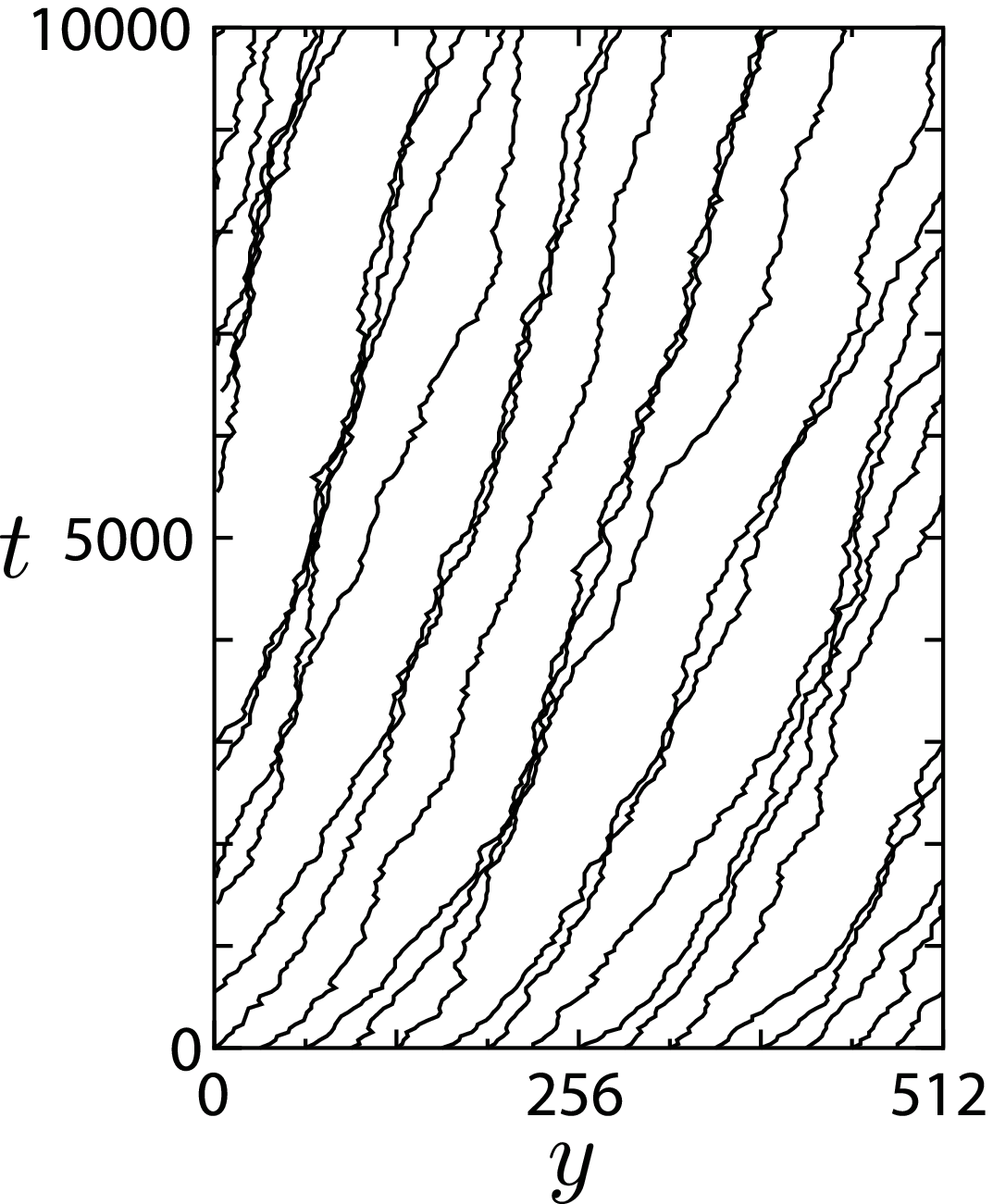}
\includegraphics[width=0.45\linewidth]{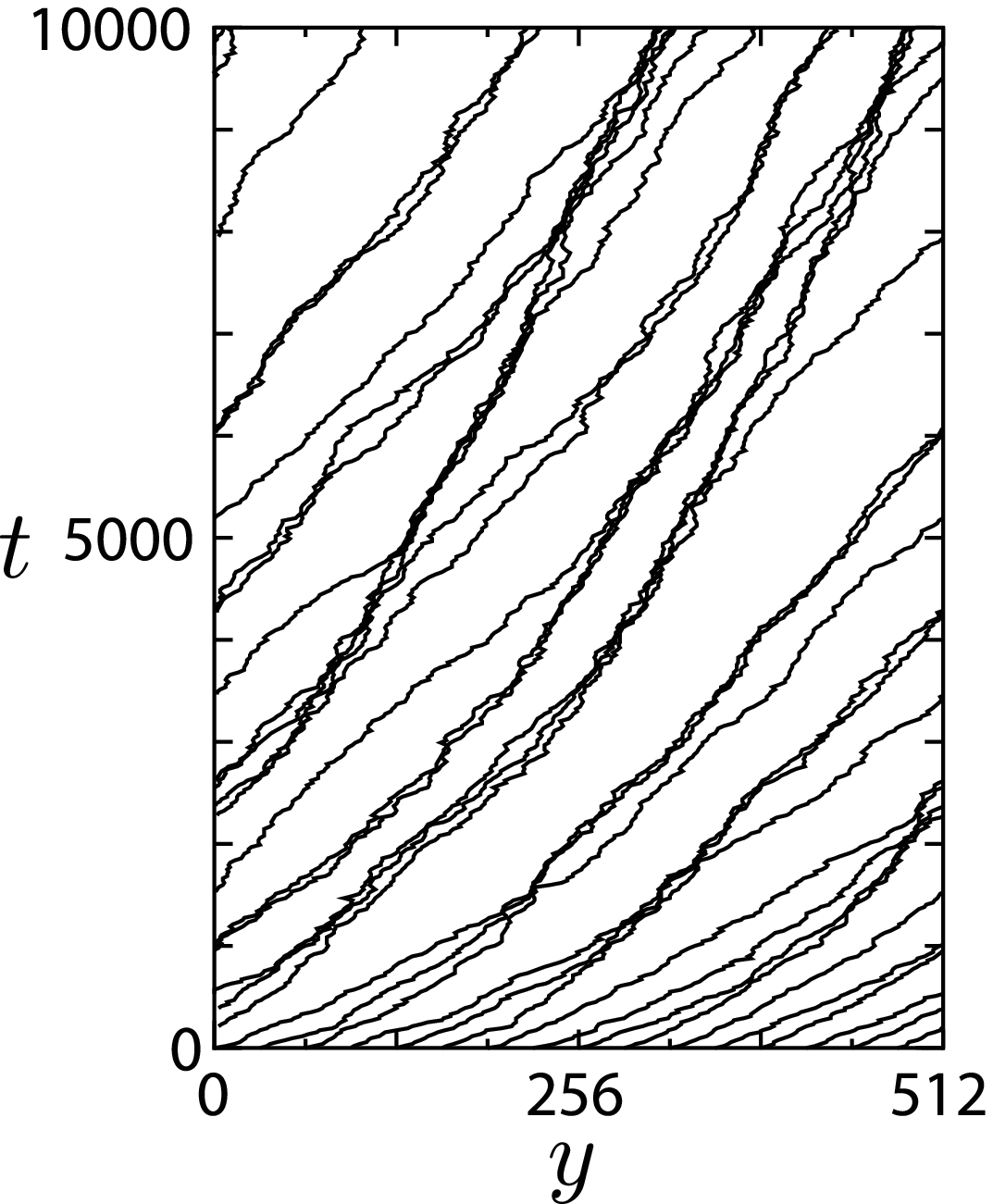}
}
\center{\Large (a) \hspace*{3.2cm} (b)}
\caption{
Time evolution of step positions with 
(a) $c_\infty=0.2$ and (b) $c_\infty=0.4$. 
The parameters are the same as those 
in Fig.~\ref{fig:time evolution of width}.
}
\label{fig:time evolution of step positions}
\end{figure}

Figure~\ref{fig:time evolution of step positions} shows 
the time evolution of step positions. 
The parameters are the same as those 
in Fig.~\ref{fig:time evolution of width}.
The value of $\alpha$ is related to the motion of steps.
In the initial stage, in which $w$ increases as $t^{1/2}$, 
single steps gather and form small bunches. 
The vicinal face separates into bunches and terraces.
Steps do not exist on the terraces. 
With increasing supersaturation,
the single steps advance rapidly
 and tight bunches are formed.
In the later stage, in which $w$ increases as $t^{1/3}$, 
both the growth of bunches by the coalescence of small bunches 
and the separation of steps repeatedly occur.

\begin{figure}[htp]
\centerline{ 
\includegraphics[width=0.5\linewidth]{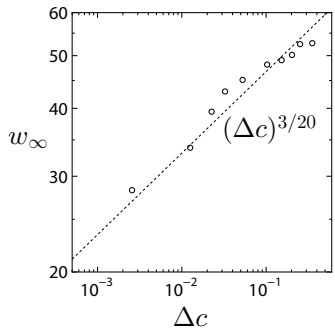}
}
\caption{
Dependence of $w_\infty$ on the supersaturation,
where $\Delta c \equiv (c_\infty-c_\mathrm{eq})$. 
The flow rate is given by $p_\mathrm{flow} =0.1$.
}
\label{fig:saturated width}
\end{figure}

In the final stage, in which the saturation of $w$ occurs,
the collision of bunches, 
the temporary breaking of bunches 
into steps and small bunches, and their recoalescence 
repeatedly occur~\cite{Sato-10jpsj}.
To determine the dependence of the saturated width $w_\infty$ on the
supersaturation, we carry out simulations with various $c_\infty$.
Figure~\ref{fig:saturated width} shows the dependence
of   $w_\infty $ on the supersaturation 
($c_\infty -c_\mathrm{eq}$).
Since we use $\phi/k_\mathrm{B}T=3.0$, 
the equilibrium adatom density is given by 
$c_\mathrm{eq} =4.7 \times 10^{-2}$.
We use the flow rate  $p_\mathrm{flow}=0.1$.
With increasing supersaturation, $w_\infty$ increases.
The dependence is given by 
$w_\infty \sim (c_\infty-c_\mathrm{eq})^\gamma$.
The value of $\gamma$ is approximately given by $\gamma =3/20$,
which is similar to the dependence of $w_\infty$ on $p_\mathrm{flow}$.

%

\section{Summary and Discussion}~\label{sec:summary}
In this paper, we carried out Monte Carlo simulations
and studied step bunching induced by a flow in a solution.
When the flow is rapid,
the width of the fluctuation of step distance increases 
with time as $t^{\alpha}$
and $\alpha $ changes with two stages.
In the first  stage, $\alpha$ is equal to $1/2$, 
which is larger than that with a slow flow.
In the second stage, $\alpha$ is $1/3$,
which is the same as that with a slow flow.
The time interval with $\alpha=1/2$ becomes longer with increasing 
supersaturation.

{
The change in the exponent is related to the motion of the steps.
When $\alpha=1/2$, bunches grow by the coalescence of single steps.
A vicinal face is separated to form a wide terrace and small bunches consisting 
a few steps.
The exponent $\alpha=1/2$ means that the correlation between steps is 
weak and that the steps move randomly.
The situation is similar to the roughing of an interface 
by ballistic deposition~\cite{barabasi}.
}
When $\alpha=1/3$, the step separation of steps from bunches repeatedly occurs. 
Single steps are present on large terraces.
In a later stage, the separation of steps is more frequent than that in an early stage. 
Owing to the single steps on large terraces, the average terrace width decreases.
Thus, the growth of the width of fluctuation becomes graduall
and $\alpha$ is lowered.
In the last stage, the saturation of the width $w$ occurs.
The saturated value $w_\infty $ increases as
$(\Delta c)^\gamma$ with $\gamma \approx 3/20$.
The dependence of $w_\infty $  on the supersaturation is the same as 
that of the flow rate~\cite{Sato-10jpsj}.
We are currently  attempting  to clarify how the exponents  can be determined.

In this paper, we assumed that the steps are straight.
Thus, step wandering, which is the instability due to the fluctuation
along steps,  cannot be treated in our model.
To study step wandering, we intend to carry out 
simulations using a two-dimensional step model.

\begin{acknowledgments}
This work was supported by a Grant-in-Aid for Scientific Research
from the Japan Society for the Promotion of Science.
\end{acknowledgments}

\end{document}